\documentclass[prl,aps,twocolumn,floatfix]{revtex4}
\usepackage{bm}
\usepackage{graphicx}
\usepackage{subfigure}

\def\be{\begin{equation}}
\def\ee{\end{equation}}
\def\bea{\begin{eqnarray}}
\def\eea{\end{eqnarray}}

\begin{document}
\def\tit#1#2#3#4#5{{#1} {\bf #2}, #3 (#4)}

\title{Flat spin wave dispersion in a triangular antiferromagnet} 
\author{Oleg A. Starykh$^1$, Andrey V. Chubukov$^2$, and Alexander G. Abanov$^3$}
\affiliation{ 
$^1$Department of Physics, University of Utah, 
Salt Lake City, UT 84112,\\
$^2$Department of Physics, University of Wisconsin, Madison, WI 53706,\\
$^3$Department of Physics and Astronomy, Stony Brook University, 
 Stony Brook, NY 11794-3800}
\date{\today}

\begin{abstract}
The excitation spectrum of a $S=1/2$ 2D triangular quantum
antiferromagnet is studied using $1/S$ expansion. Due to the non-collinearity of the classical 
ground state significant and non-trivial corrections to the spin wave spectrum appear already in 
the first order in $1/S$ in contrast to the square lattice antiferromagnet. The resulting magnon dispersion 
is almost flat in a substantial portion of the Brillouin zone. Our results are in quantitative agreement with 
recent series expansion studies by Zheng, Fj\ae restad, Singh, McKenzie, and Coldea 
[\prl {\bf 96}, 057201 (2006) and cond-mat/0608008].
\end{abstract}

\maketitle

Triangular antiferromagnets occupy a special niche in the  
studies of quantum magnetism.
Ising antiferromagnet on triangular lattice has a finite 
zero-temperature entropy, which reflects an extensive degeneracy of the 
ground state manifold \cite{wannier}. Classical Heisenberg
model on a triangular
 lattice represents the textbook example of the full $SU(2)$ 
 symmetry breaking and noncollinear 
{\em spiral} spin ordering in the 
ground state - the $120^\circ$ spin structure. 
For a quantum $S=1/2$ antiferromagnet on a triangular lattice, 
Anderson proposed back in 1973 the disordered {\em resonating valence bond} (RVB) ground state \cite{rvb}. 
This suggestion stimulated extensive research for over 25 years. An
 RVB ground state  on a
 triangular lattice has been found recently \cite{qdm-rvb}, albeit for a quantum dimer model. It was also established, by  large-N and gauge theory approaches,
 that a disordered ground state of a triangular antiferromagnet
 must possess unconfined  massive spinon excitations \cite{read-sachdev}.   

On the experimental side, several novel materials with triangular structure
have attracted substantial interest over the last few years. 
 $Na_xCoO_2$, 
in which $Co$ atoms form a layered hexagonal structure~\cite{nacoo},
was suggested to be in close
 proximity to a spin liquid~\cite{lee}.  Another potential candidate for a  spin liquid is 
$\kappa-(ET)_2Cu_2(CN)_3$ at small pressures \cite{kanoda}.  Finally, there is  
an intensive theoretical debate~\cite{veillette1,veillette2,dalidovich,alicea} 
on the structure of the ground state of  a spatially anisotropic
 triangular $S=1/2$ antiferromagnet $Cs_2CuCl_4$ \cite{coldea2001}. 

The ideas about the disordered ground state of unconfined spinons, 
however, could {\em not} be
 immediately applied to the  most studied  Heisenberg model of 
 quantum $S=1/2$ spins 
 on a triangular lattice,
 as both perturbative $1/S$ and numerical calculations point that the classical, $120^\circ$ spin 
structure survives quantum fluctuations.
Quantum fluctuations do reduce the average value of a sublattice magnetization
 to $50\%$ of its classical value~\cite{css}. 
 This renormalization is  
 generally comparable to that
 for a $S=1/2$ antiferromagnet on the square lattice~\cite{igarashi}.
 For the latter, calculations
 to order $1/S^2$ for the spectrum do show that the
overall scale of the spin-wave dispersion is 
 renormalized by quantum fluctuations, but the dispersion
 retains almost the same functional form as in the quasiclassical limit, 
 and obviously is better described by  magnons rather than by deconfined spinons.
 
For a triangular antiferromagnet, 
Chubukov et al~\cite{css} computed $1/S$ corrections to the two spin-wave velocities, and found that these corrections are quite small, even for $S=1/2$.   
No $1/S$ calculations of the full spin-wave
 dispersion have been reported in the literature but, based on the results for the velocities, it was 
 widely believed that  the functional form of the
dispersion for $S=1/2$ 
antiferromagnet should also be close to that in the quasiclassical, 
large $S$ limit, i.e., that the spin-wave description can be extended to $S=1/2$.    
 Recent series expansion study~\cite{rajiv_1}, however, uncovered 
remarkable changes in the functional form of the
 dispersion in the isotropic quantum $S=1/2$ antiferromagnet 
on triangular lattice compared to $S=\infty$ limit. In particular, the dispersion for the $S=1/2$ case possesses
 local minima (``rotons") at the mid-points of faces of the hexagonal Brillouin zone (BZ). The classical dispersion does not have such local minima. 

The authors of Ref. \cite{rajiv_1} conjectured that the qualitative changes between the actual dispersion for $S=1/2$ and the classical dispersion may imply
 that at energies comparable to the exchange integral $J$, the system is better
 described in terms of pairs of deconfined spinons rather than magnons
 (the latter in such description are bound states of spinons). In other words,
 they argued that the spinon  description, valid for the disordered state,
 may adequately describe high-energy excitations of the ordered state.

 In this communication, we propose another explanation for the series expansion  results, 
alternative to the one proposed in   Ref. \cite{rajiv_1}.
   We argue that  regular 1/S corrections, extended to $S=1/2$,
strongly modify the form of the magnon dispersion in a triangular antiferromagnet, and the renormalized dispersion has 
``roton" minima at the faces of the BZ, in agreement with~\cite{rajiv_1}.
 But we also found a more drastic effect -- 
the renormalized dispersion turns out to 
 be almost flat in a wide range of momenta.
The flat renormalized dispersion has not been 
 reported in~\cite{rajiv_1} where the numerical data have been presented only along special high-symmetry  directions in the BZ. 
The subsequent, more detailed
 series expansion studies~\cite{rajiv_2}, carried out in parallel 
with our research, did find the regions of flat dispersion. 
The analytical and series expansion results are in
 good qualitative and quantitative agreement, which 
 indicates that the flat regions are likely present in the actual dispersion,
 and  that  a first order in $1/S$ provides fairly accurate description of a 
 triangular antiferromagnet even at high energies, comparable to the exchange $J$.

\begin{figure}
  \centering
  \includegraphics[width=3.0cm]{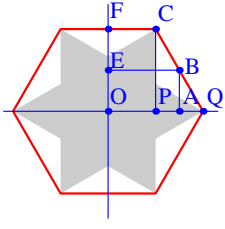}
  \caption{(Color online) BZ of a triangular antiferromagnet. 
Thin lines indicate the directions along which the ``cuts" of dispersion are taken. 
Coordinates of the points are: P$(2\pi/3,0)$, A$(\pi,0)$, 
Q$(4\pi/3,0)$, B$(\pi,\pi/\sqrt{3})$, 
C$(2\pi/3,2\pi/\sqrt{3})$, F$(0,2\pi/\sqrt{3})$, 
and E$(0,\pi/\sqrt{3})$. ${\text{Im}}\Sigma^{(S)}_k \neq 0$ inside the
shaded star-shaped area.}    
\label{fig:bz}
\end{figure}

The point of departure for the $1/S$ calculation is the
 Heisenberg antiferromagnet on a triangular lattice 
\be 
{\cal H} = J \sum_{<i,j>} {\bf S}_i {\bf S}_j .
\label{1}
\ee
Using Holstein-Primakoff transformation to bosons, assuming
the $120^\circ$ spin structure in the ground state, and diagonalizing the 
 quadratic form in bosons one can  
 re-write Eq. (\ref{1}) as \cite{css}
\be
{\cal H} = 3 JS \left[\sum_k E_k c^\dagger_k c_k  + \frac{1}{\sqrt{S}} 
{\cal H}_3 + \frac{1}{S} {\cal H}_4 + ...\right],
\label{2}
\ee
where dots stand for higher order terms in $1/S$, and 
\be
E_k = \sqrt{A^2_k - B^2_k} = \sqrt{(1-\nu_k)(1+2\nu_k)},
\label{3_1}
\ee
where 
\bea
A_k &=& 1 + \frac{\nu_k}{2}, ~B_k = -\frac{3}{2} \nu_k, \nonumber\\
\nu_k &=& \frac{1}{3} \left(\cos{k_x} + 2 \cos {\frac{k_x}{2}} \cos{\frac{\sqrt{3}k_y}{2}} \right).
\label{3}
\eea
The BZ is presented in Fig.~\ref{fig:bz}. The magnon energy $E_k$ vanishes at the center of the zone, $k=0$, where $\nu_k = 1$, and at the corners of the BZ (e.g. at points Q and C), where $\nu_k = -1/2$.

In a square-lattice antiferromagnet, ${\cal H}_3$ term is absent, and 
 $1/S$ corrections to the dispersion come from the decoupling  of the four-boson term. These corrections do not affect the functional form of the dispersion~\cite{igarashi}. The corrections to the dispersion then appear at $1/S^2$ order, from ${\cal H}_6$ term in the magnon Hamiltonian and from the second-order perturbation in ${\cal H}_4$,
both are very small numerically. In triangular antiferromagnets, ${\cal H}_3$ term is present due to the {\em non-collinearity} of the classical spin configuration
 ($120^\circ$ structure) and, taken at the second order, 
 gives rise to non-trivial corrections to the dispersion already at the order $1/S$.
The expressions for ${\cal H}_3$ and ${\cal H}_4$ have been obtained in \cite{css} and we refer to that work for the derivation.
 Decoupling the four-magnon term in a standard way and adding the result 
 to the quadratic form we obtain, to order $1/S$, 
\be
{\cal H} = 3 JS \left[\sum_k {\bar E}_k c^\dagger_k c_k  +
 \frac{1}{\sqrt{S}} {\cal  H}_3\right] ,
\label{4}
\ee
where ${\bar E}_k$ represents dispersion renormalized by quartic terms (coming from
${\cal H}_4$ \cite{css}) 
\be
{\bar E}_k^{2} =  E_k^2 (1+c_1/S) + (1-\nu_k)c_2/S ,
\label{5}
\ee
where $c_1=1-I_0 - I_1/4 + 5I_2/4 ,~c_2=3(I_2-I_1)/4$
and $I_{n} = \frac{1}{N}\sum_k \nu_k^n/E_k$. The cubic part reads
\bea
&&{\cal  H}_3 =  i\left(\frac{3}{32 N}\right)^{1/2} 
\sum_{1,2} \Big( c^\dagger_1 c^\dagger_2 c_{k} \Phi_1 (1,2,k) \delta_{1+2-k} 
\nonumber \\
&& + \frac{1}{3} c^\dagger_1 c^\dagger_2 c^\dagger_{k} \Phi_2 (1,2,k) 
\delta_{1+2+k} + {\text{h.c.}} \Big) .
\eea
Here $1 = k_1$, $2 = k_2$, and summation is over the BZ.
The vertices  $\Phi_{1,2} (1,2,k) = {\tilde \Phi}_{1,2} (1,2,k)/\sqrt{E_1 E_2 E_k}$, where 
\bea  
&&{\tilde \Phi}_{1,2} (1,2,3)  = {\bar \nu}_1 f^{(1)}_{-} \left( f^{(2)}_{+} f^{(3)}_{+}
\pm  f^{(2)}_{-} f^{(3)}_{-}\right) + {\bar \nu}_2 f^{(2)}_{-} \times \nonumber \\
&& \left( f^{(1)}_{+} f^{(3)}_{+}
\pm  f^{(1)}_{-} f^{(3)}_{-}\right) + {\bar \nu}_3 f^{(3)}_{-} \left( f^{(1)}_{+} f^{(2)}_{+} -
  f^{(1)}_{-} f^{(2)}_{-}\right)
\label{7}
\eea
are expressed in terms of $f^{(i)}_{\pm} = \sqrt{A_{k_i} \pm B_{k_i}}$ and
\be
{\bar \nu}_k = \frac{2}{3} \sin \frac{k_x}{2} \left(\cos \frac{k_x}{2} - \cos \frac{\sqrt{3}k_y}{2}\right).
\label{8}
\ee  
The ${\cal  H}_3$ term gives rise to a $k$-dependent magnon self-energy to order $1/S$
\bea
\Sigma_k^{(S)} &=& -\frac{3}{16 SN} \Big( \sum_{1+2 =k} 
\frac{|\Phi_1 (1,2,k)|^2}{E_1 + E_2 - E_k + i 0}  \nonumber\\
&& + \sum_{1+2 =-k} \frac{|\Phi_2 (1,2,k)|^2}{E_1 + E_2 + E_k + i 0}\Big) .
\label{9}
\eea
Note that at this order $\Phi_{1,2}$ and $E_k$ are expressed via bare ($S=\infty$)
quantities. Collecting the corrections from  three-magnon and four-magnon 
 processes and restoring the pre-factor $3JS$ (see Eq.~\ref{4}), we obtain for the full renormalized dispersion
\be
E_{\text{ren}} (k) = 3JS \sqrt{{\bar E}_k^2 + 2 E_k \Sigma_k^{(S)}} .
\label{10}
\ee
Expanding (\ref{10}) to first order in $1/S$ we finally obtain
\be\tighten
E_{\text{ren}} (k) = 3 J S \left[E_k (1 + \frac{c_1}{2S}) + 
\frac{(1-\nu_k) c_2}{2S E_k} + \Sigma_k^{(S)}\right] .
\label{11}
\ee

It was shown in \cite{css} that the renormalized dispersion preserves zeros 
at $k=0$ and at the corners of the BZ, i.e., locations of Goldstone modes are 
not affected by $1/S$ corrections. 
Below we compute the full spin-wave dispersion (\ref{11}) 
and apply the results to $S=1/2$.

We used the Mathematica{\copyright}  software to calculate the self-energy $\Sigma_k$, 
(\ref{9}).
Two-dimensional momentum integrals were regularized by replacing energy
denominators with $E+i\delta$, where $\delta = 10^{-4}$,
and taking the real part of the resulting expression. The imaginary part was 
calculated using the ``bell approximation" for the delta function: 
$\delta(x) = \sqrt{t/\pi} e^{-t x^2}$. We found that $t=10^3$ gives
 stable and consistent results.

\begin{figure}
\centering
\subfigure[ ] 
{
    \label{fig:3d-a}
    \includegraphics[width=4cm]{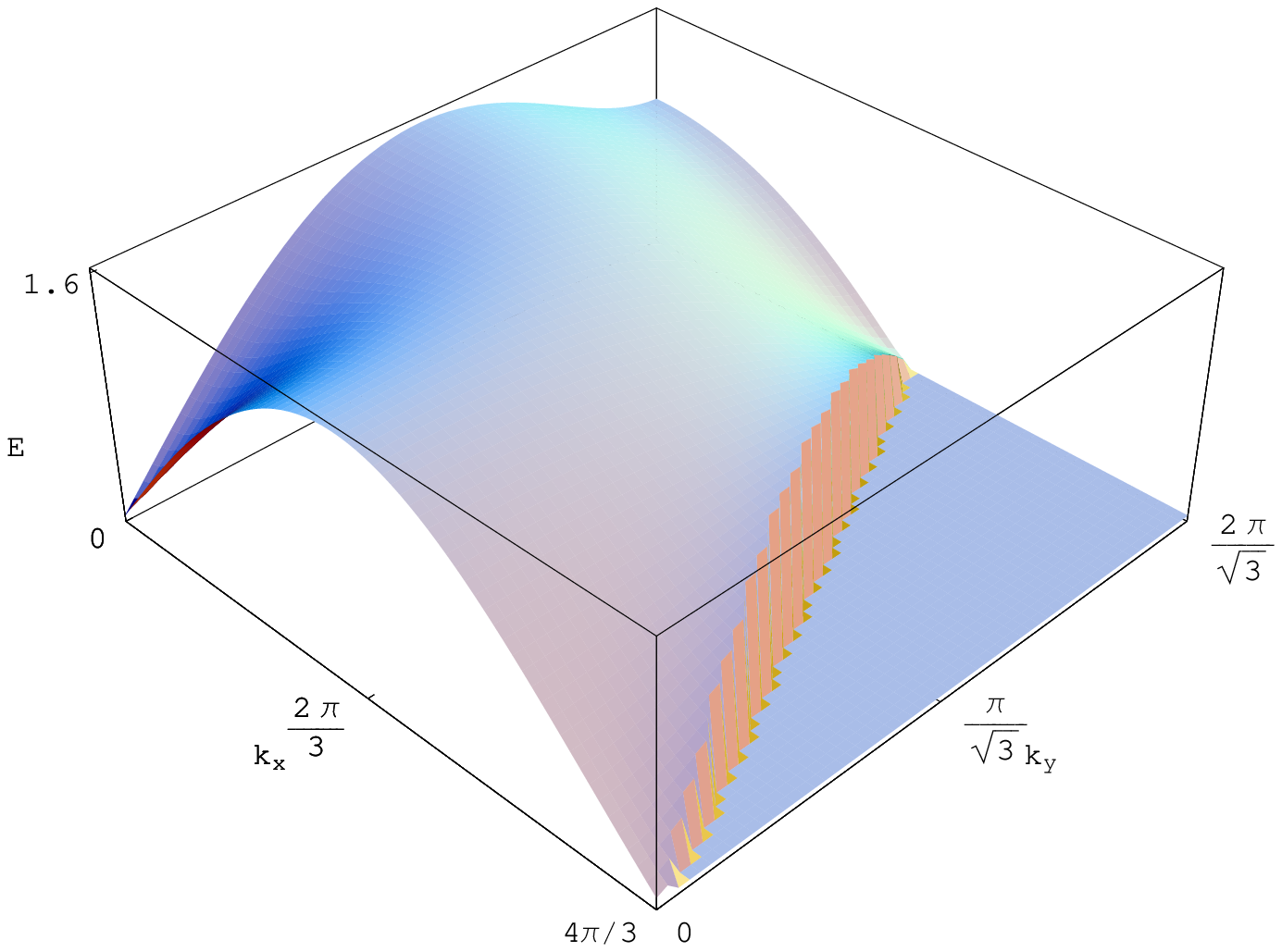}
}
\hspace{0cm}
\subfigure[ ] 
{
    \label{fig:3d-b}
    \includegraphics[width=4cm]{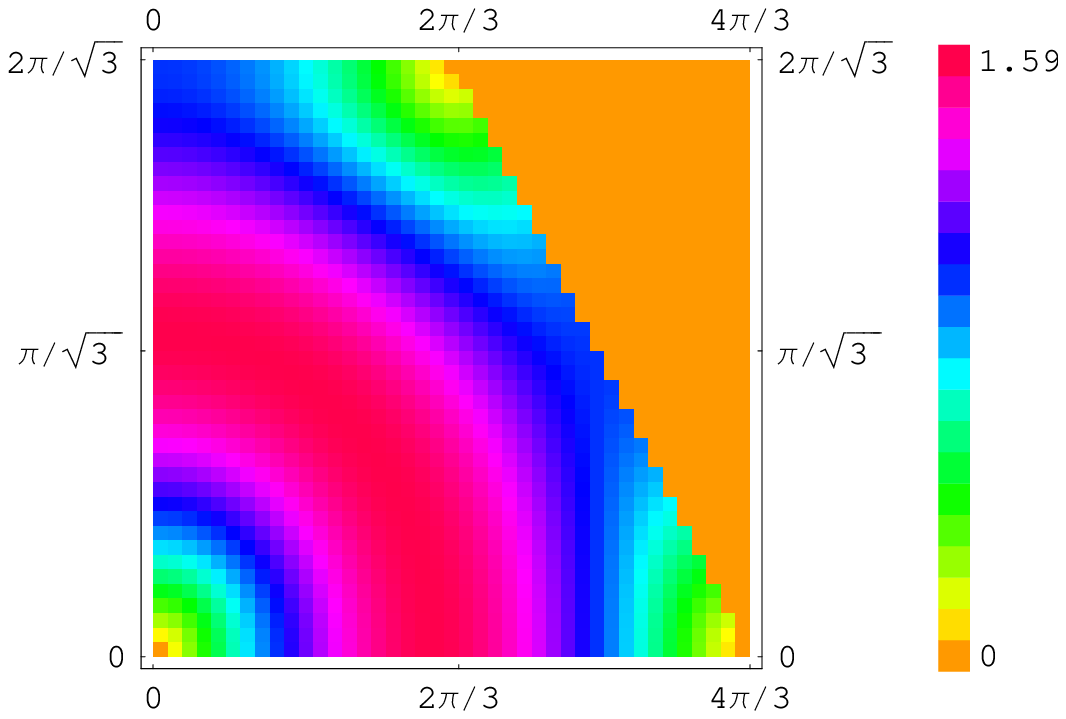}
}
\vspace{-.2cm}
\subfigure[ ] 
{
    \label{fig:3d-c}
    \includegraphics[width=4cm]{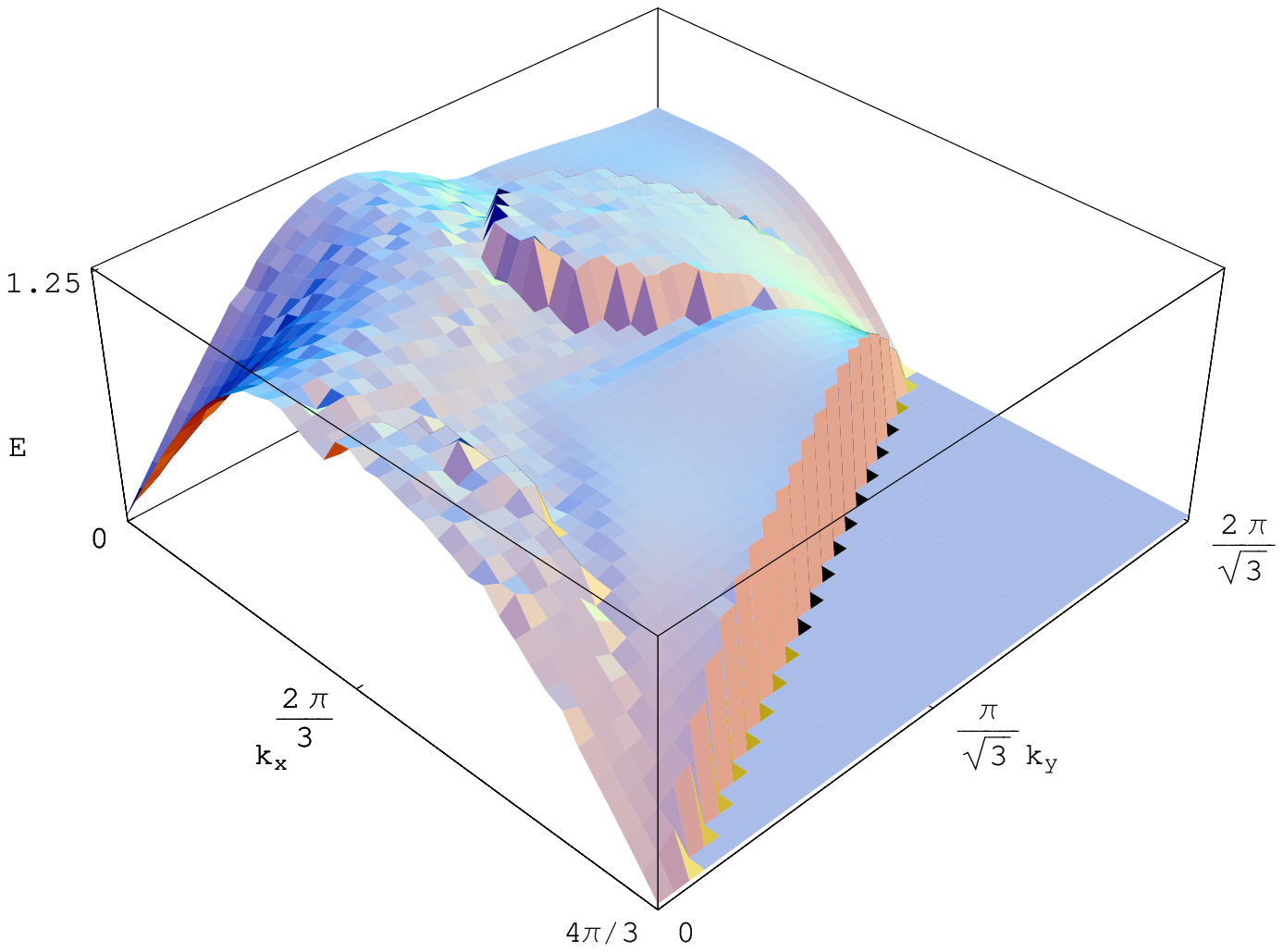}
}
\hspace{0cm}
\subfigure[ ] 
{
    \label{fig:3d-d}
    \includegraphics[width=4cm]{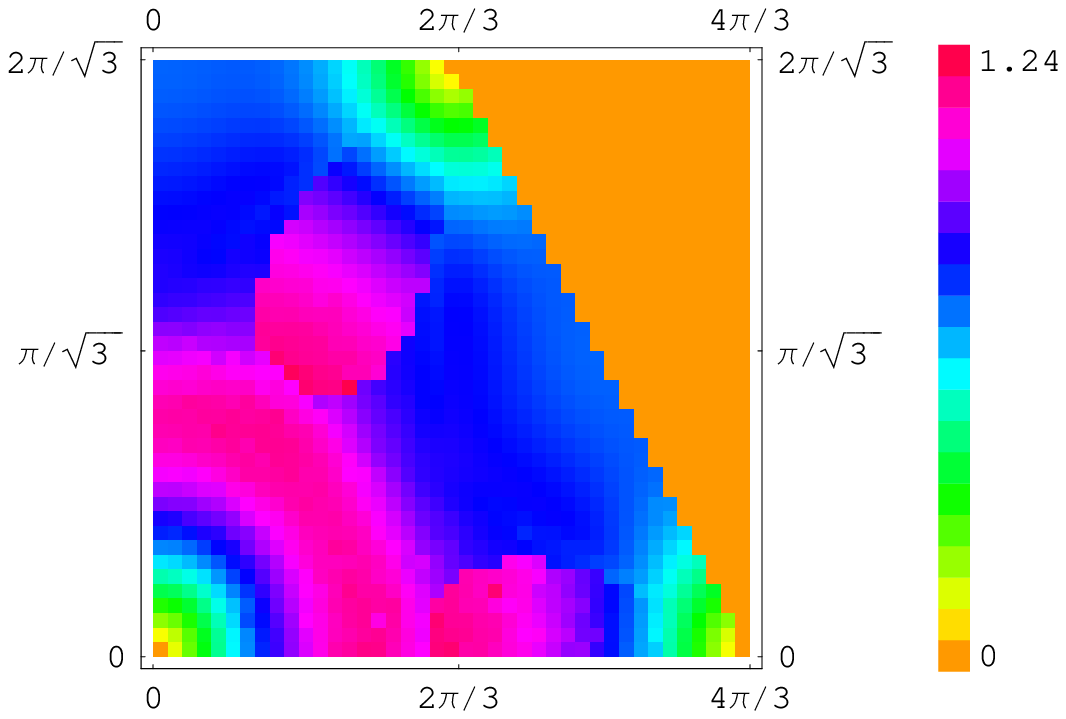}
}
\vspace{-.2cm}
\subfigure[ ] 
{
    \label{fig:3d-e}
    \includegraphics[width=4cm]{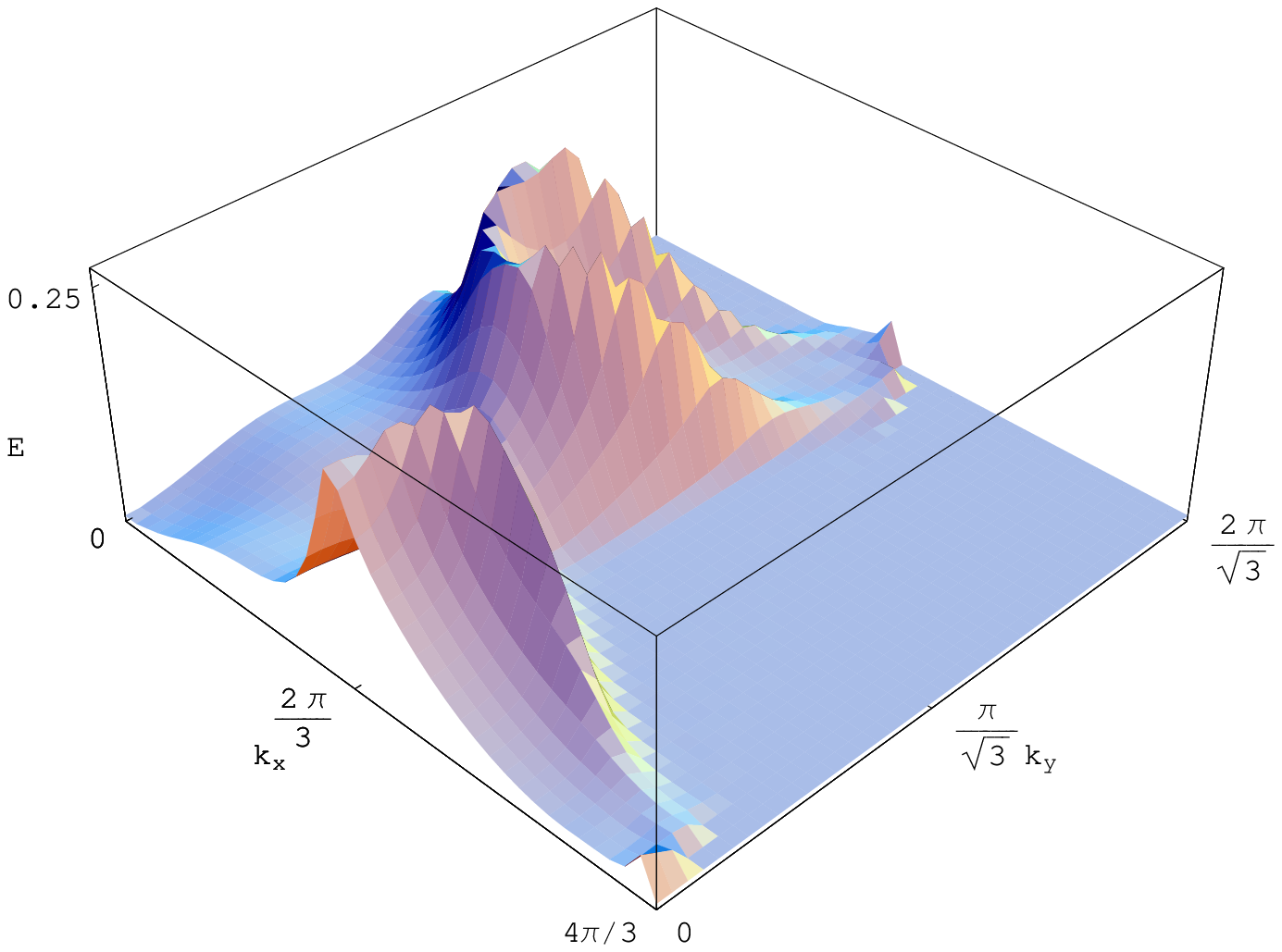}
}
\hspace{0cm}
\subfigure[ ] 
{
    \label{fig:3d-f}
    \includegraphics[width=4cm]{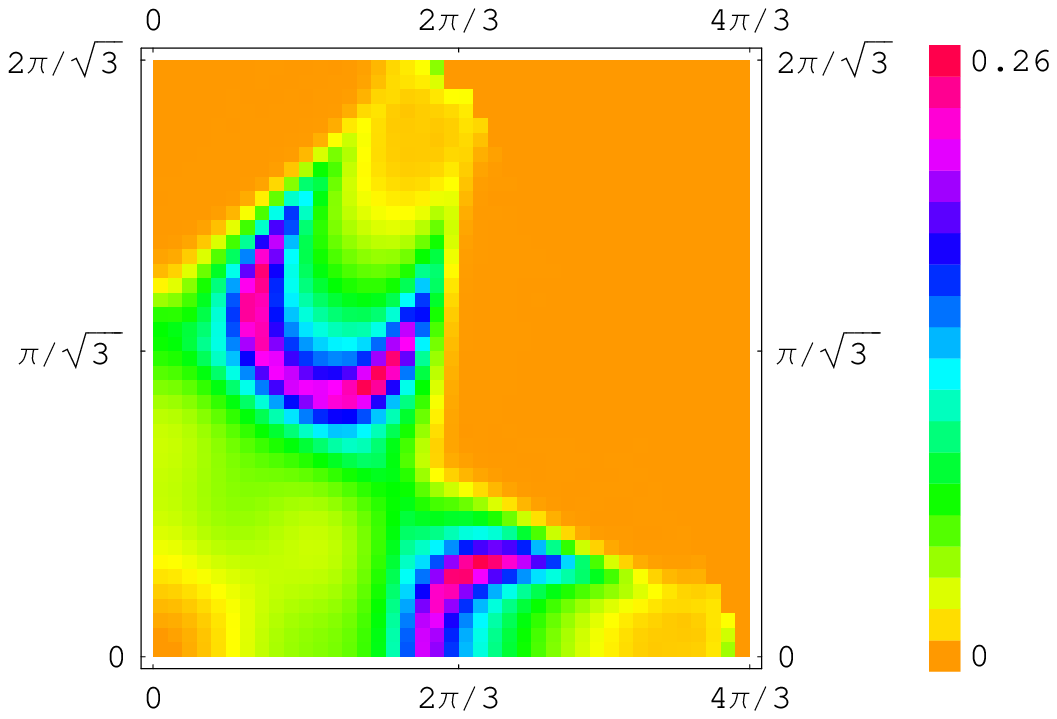}
}
\caption{(Color online) 
Classical  and  renormalized  spin wave dispersions.
Left panels -- three-dimensional plots, right panels -- color density plots.
The dispersions are shown in the OQCF quadrant of the BZ and are
 set to zero outside this region to ease the viewing. 
The classical dispersion, figures (a,b), is $1.5 E_k$. 
The real part of
 renormalized dispersion  $E_{\text{ren}} (k)$ (Eq. \ref{11}) is shown in figures (c) 
 and (d). The flat dispersion is in the blue (dark) region in (d).
 The imaginary part of the 
renormalized dispersion is shown in  figures (e, f). Observe that it is numerically small 
throughout the BZ, and is zero outside the ``star'' region, see also Fig~\ref{fig:bz}. }
\label{fig:3d} 
\end{figure}

Our results are
shown as three-dimensional plots in Fig~\ref{fig:3d}. For comparison we also plotted 
the classical  dispersion $3JS E_k = 1.5J E_k$. 
For clarity, we only plot the dispersion over a quarter of the BZ, 
and set it to zero outside this quarter.  Observe that the renormalized 
dispersion vanishes at 
$k=0$ and at the corners of the BZ hexagon. 
Spin wave velocity 
at $k=0$ decreases in comparison to the classical result, while the one
at the corners of the  BZ  increases
(see Fig~\ref{fig:OQ} below)~\cite{css}.
 
\begin{figure}
\centering
\subfigure[ ] 
{
    \label{fig:FC}
    \includegraphics[width=4cm]{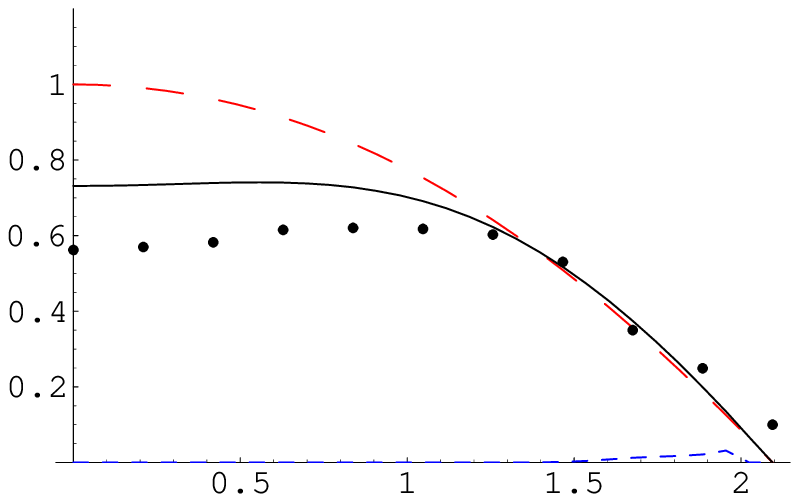}
}
\hspace{0cm}
\subfigure[ ] 
{
    \label{fig:OF}
    \includegraphics[width=4cm]{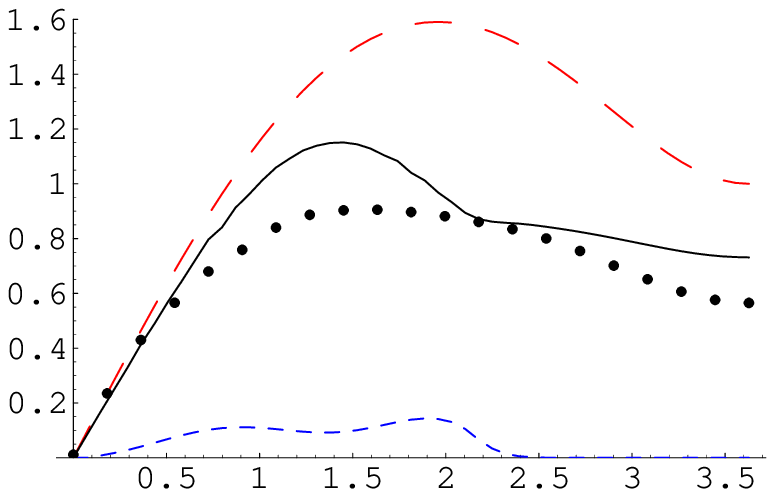}
}
\caption{(Color online) Classical  and renormalized dispersions (ordinate) 
along (a) FC and (b) OF directions (abscissa). 
Red dashed line - classical dispersion. Black solid line --
 the real part of the renormalized dispersion $E_{\text{ren}} (k)$, Eq.~\protect(\ref{11}).
 Blue dotted line - the imaginary part of 
$E_{\text{ren}} (k)$.
The black dots are series expansion data from \cite{rajiv_2}.
No fitting parameter is involved in the comparison 
 with the data.}
\label{fig:FC-OF} 
\end{figure}

We clearly see from Fig~\ref{fig:3d} that the actual 
dispersion is rather different from $1.5 E_k$. The key difference is
 that the renormalized dispersion $E_{\text{ren}} (k)$
 has a plateau at around $0.8 J$ 
over a wide range of momenta. This is most clearly seen in Fig ~\ref{fig:3d-c} and \ref{fig:3d-d}. 
The dispersion also possesses a roton-like minimum near 
 the faces of the BZ, as is best seen along the cut FC, Fig~\ref{fig:FC}.

In Fig~\ref{fig:3d-e} and \ref{fig:3d-f} we show the imaginary part of 
$E_{\text{ren}} (k)$ from three-magnon processes.  In distinction to a 
square-lattice antiferromagnet, the imaginary part of the dispersion 
in our case appears already at the order $1/S$. 
As follows from (\ref{9}), it is present when one-particle excitation (magnon) 
with momentum $\vec{k}$ can decay into the two-particle continuum, i.e., when 
$E_{\vec{q}} + E_{\vec{k} - \vec{q}} = E_{\vec{k}}$ for some $\vec{q}\in$ BZ.
This condition defines star-shaped region, shown in 
light gray shading in Fig~\ref{fig:bz} (see also Fig~\ref{fig:3d-f}): inside it 
${\text{Im}}\Sigma^{(S)}_k$ is nonzero.  
While ${\text{Im}}\Sigma^{(S)}_k \neq 0$ 
(and, hence, ${\text{Im}}E_{\text{ren}} (k) \neq 0$)
in most of the BZ, we found that
${\text{Im}}E_{\text{ren}} (k)$ 
does not exceed  $0.26 J$, and  
is much smaller than ${\text{Re}}E_{\text{ren}} (k)$. Nonetheless, a finite
imaginary part is important as it responsible for dampening excitations at wavevectors where variations of ${\text{Re}}E_{\text{ren}} (k)$ in Fig~\ref{fig:3d-c} and \ref{fig:3d-d} are maximal.  

Observe that the  maxima of ${\text{Im}}E_{\text{ren}} (k)$ in Fig~\ref{fig:3d-e}
 occur exactly where the variation of ${\text{Re}}E_{\text{ren}} (k)$ is the strongest,
 see also Figs~\ref{fig:FC-OF}-\ref{fig:AB-EB} below.  
In other words, the sharp features in the $1/S$ dispersion are 
  not artefacts of numerical calculations, but are real features of the 
 excitation spectrum at this order. 
Observe also that  ${\text{Im}}E_{\text{ren}} (k)$ vanishes at momenta where 
 ${\text{Re}}E_{\text{ren}} (k)$ stays
almost constant. This implies that nearly immobile magnons 
have infinite lifetime, i.e., are true excited states of the system. 
\begin{figure}
\centering
\subfigure[ ] 
{
    \label{fig:OQ}
    \includegraphics[width=4cm]{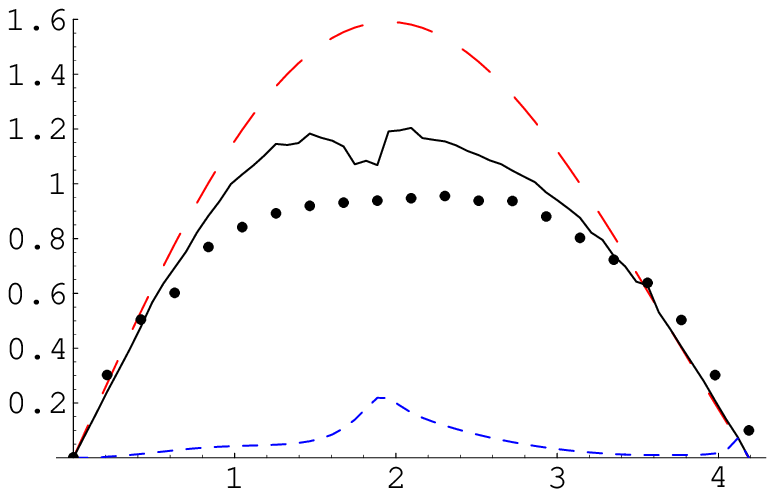}
}
\hspace{0cm}
\subfigure[ ] 
{
    \label{fig:PC}
    \includegraphics[width=4cm]{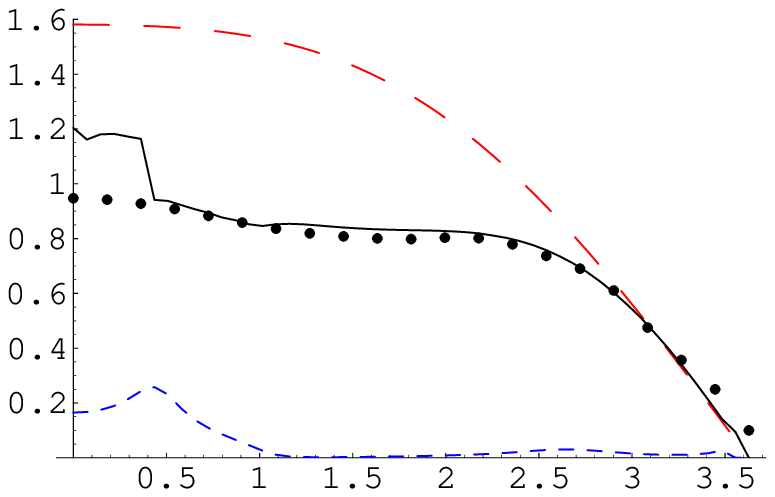}
}
\caption{(Color online) Same as in Fig.~\ref{fig:FC-OF} but  along (a) OQ and (b) PC directions.}
\label{fig:OQ-PC} 
\end{figure} \begin{figure}
\centering
\subfigure[ ] 
{
    \label{fig:AB}
    \includegraphics[width=4cm]{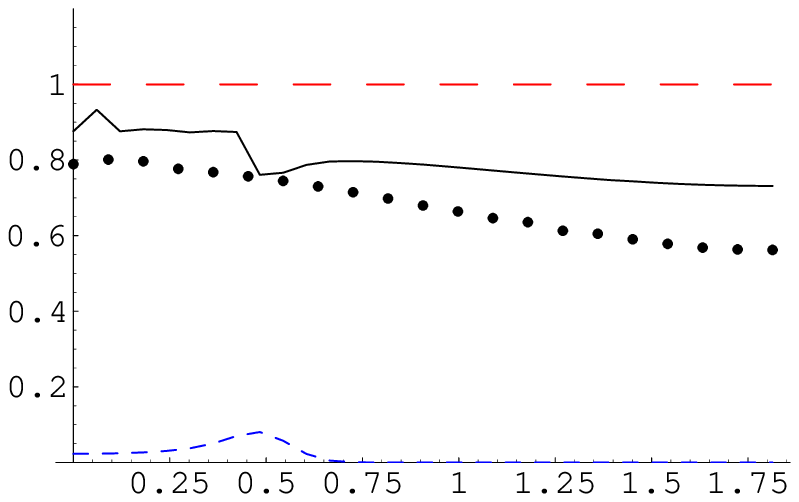}
}
\hspace{0cm}
\subfigure[ ] 
{
    \label{fig:EB}
    \includegraphics[width=4cm]{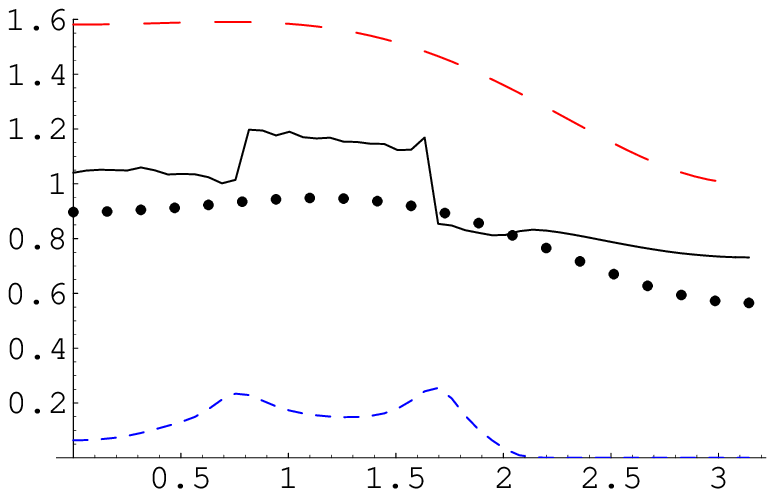}
}
\caption{(Color online) Same as in Fig.~\ref{fig:FC-OF} but 
along (a) AB and (b) EB directions.}
\label{fig:AB-EB} 
\end{figure}

To better display the new features and for comparison with series expansion studies, we show in Figs~\ref{fig:FC-OF}-\ref{fig:AB-EB} the dispersion $E_{\text{ren}} (k)$ along six different cuts through the BZ (black lines). The directions of
 particular cuts are shown in Fig~\ref{fig:bz}. In each of the plots, 
we also show the classical dispersion (dashed red lines). The 
flat regions are clearly seen in the cuts along FC, OF, PC, and EB directions, 
and the roton minimum is seen in the cut along FC direction.
Note that in accordance with Fig~\ref{fig:bz}, the local minimum near point F 
(Fig~\ref{fig:FC})
corresponds to a truly stable excitation, ${\text{Im}}E_{\text{ren}} (k) = 0$ there.
The maximum of the renormalized dispersion 
is at $1.2 J$ which is substantially
smaller than $1.6 J$ for the classical dispersion.  

We now compare in some detail 
our results to series expansion studies of Zheng et al 
\cite{rajiv_1,rajiv_2}.  First, series expansion
 studies have found that the dispersion has a 
bandwidth of about $J$ (see series data in Fig~\ref{fig:PC}),  which is 
 substantially smaller than $1.6J$ for a classical dispersion. 
This is in agreement with our results 
(see Figs.~\ref{fig:FC-OF}-\ref{fig:AB-EB}). 
 Note that this difference is much larger than one might expect by comparing the spin-wave velocities which are renormalized only by about 10\%~\cite{css}.    
 Second, series expansion studies found ``roton" minima near points F 
and B in the BZ, Fig~\ref{fig:bz}. Our cuts along FC and OF 
also  show a minumum near point F (see Fig~\ref{fig:FC-OF}).
Third, recent series expansion studies~\cite{rajiv_2}
 found the regions of flat dispersion at $0.8J$ 
along FC, PC, and EB directions. Our results for 
$E_{\text{ren}} (k)$ also show the regions of flat dispersion at around 
$0.8-0.9 J$. The flat regions are clearly seen on the density plot in 
Fig~\ref{fig:3d-d}, as well as in Figs~\ref{fig:FC},
~\ref{fig:PC}, and in Fig~\ref{fig:EB} (near points E and B). 

The near-flat regions of the magnon dispersion have a large density of states 
 and can be probed by Raman scattering.
The two-magnon Raman intensity in a square-lattice antiferromagnet 
is peaked slightly below  twice the frequency at which 
the magnon density of states diverges. In our case, the density of states is peaked at around $0.8J$, and we expect that the two-magnon 
Raman intensity will be peaked somewhat below $1.6J$. 

To conclude, in this paper we used $1/S$ expansion, extended it to $S=1/2$, 
 and obtained the renormalized magnon dispersion 
 for  a Heisenberg antiferromagnet on triangular lattice. We found that 
 the renormalized dispersion is qualitatively different from the classical one -- 
 it is almost flat over a wide range of momenta in the BZ,
 and has roton-like minuma near  the mid-points of faces of the BZ. 
 These results are in full agreement with recent series expansion studies. 

We are thankful to R.R.P. Singh for valuable discussion and to 
R.R.P. Singh and W. Zheng for sending us series expansion results prior to publication.
OAS acknowledges the donors of the American Chemical Society Petroleum Research
Fund for support (PRF43219-AC10).
AVC and AGA are supported by NSF via grants No. DMR-0240238
and DMR-0348358, respectively.
The authors thank the 
Theory Institute at Brookhaven National Laboratory where
this work has initiated.
 
{\sl Note added:} after submission of our manuscript we learned of a closely related study
of magnon decays in triangular antiferromagnets \cite{sasha-misha}.

\end{document}